# Emergent interweaved CDW unoccupied states in hole-doping $LaTe_2$ with element substitution

Shuya Xing(邢淑雅) [1,2,3,+,*], Hongyu Liu(刘宏宇)[1,2,3,+], Xu Chen(陈旭)[6], Zhenkai Xie(谢圳楷)[7], Zhongxu Wei(魏忠旭)[6], Shifeng Zhao(赵世峰)[1,2,3], Wenping Zhou(周文平)[1,2,3,*], Xin qi Li(李新奇)[1,2,3] and Zhihai Cheng(程志海)[4,5*]

[1] *School of Physical Science and Technology, Inner Mongolia University, Hohhot 010021, China*

[2] *Research Center for Quantum Physics and Technologies, Inner Mongolia University, Hohhot 010021, China*

[3] *Inner Mongolia Key Laboratory of Microscale Physics and Atom Innovation, Inner Mongolia University, Hohhot 010021, China*

[4] *Key Laboratory of Quantum State Construction and Manipulation (Ministry of Education), Renmin University of China, Beijing, 100872, China*

[5] *Beijing Key Laboratory of Optoelectronic Functional Materials & Micro-Nano Devices, School of Physics, Renmin University of China, Beijing 100872, China*

[6] *Beijing National Laboratory for Condensed Matter Physics, Institute of Physics, Chinese Academy of Sciences, Beijing 100190, China*

[7] *Department of Physics, Southern University of Science and Technology, Shenzhen, Guangdong 523803, China*

**Abstract:** Multiple CDW-ordered layered rare-earth tellurides have increasingly emerged as a research hotspot, owing to their unconventional CDW formation, high transition temperature, and confirmed existence of axial Higgs modes. Recently, interweaved CDW in $LaTe_2$ and its element-substituted phase $LaTe_{2-x}Sb_x$ have been investigated through TEM and ARPES measurements, revealing their distinct origins. Nevertheless, several complex diffraction features observed in TEM patterns remain unelucidated. In this work, we carried out scanning tunneling microscopy (STM) on $LaTe_{1.6}Sb_{0.4}$ crystals at 9 K. Three interweaved CDW wave vectors, $q_1$=8/11$a^*$, $q_2$=5/11$a^*$and $q_3$=3/11$a^*$ were observed, which are induced by hole doping in unoccupied states. The $q_1$ and $q_3$ are theoretically verified to be nesting vectors connecting $p_x$- and $p_y$- bands. Furthermore, the satellite spots relative to the main Bragg spots $p$ corresponding to a 11-$a$-superlattice have also been detected. Our findings provide critical insights for further exploring the origin of the interweaved CDW in hole/electron doping materials.

[+] These authors contributed equally: Shuya Xing and Hongyu Liu

*To whom correspondence should be addressed: 111992026@imu.edu.cn pywpzhou@imu.edu.cn zhihaicheng@ruc.edu.cn

## Introduction

Rare earth tellurides $RTe_2$ (R=rare earth element) are a class of layered materials that exhibit high CDW transition temperatures and pressure-induced superconductivity[1]. The structure of $RTe_2$ crystal adopts the layered tetragonal $Cu_2Sb$-type structure, in which a Te-Te square sheet is sandwiched between the corrugated double R-Te slabs, with the layers stacked alternately along the *c*-axis. Different from the conventional charge density wave materials[2-6] such as transition metal dichalcogenides (TMDCs) 1T-$TaS_2$, 1T-$TaSe_2$, and 2H-$NbSe_2$, $RTe_2$ shows an unconventional CDW characteristic due to its 5*p* orbitals-defined Fermi surface structure of Te anions[7].Furthermore, theoretical predictions suggest that two-dimensional layered rare earth tellurides possess hidden one-dimensional characteristics due to the anisotropic nature of the *p*-orbital electron wavefunctions. At different temperatures, they can exhibit a unidirectional "striped" or a bidirectional "checkerboard" CDW state; experimentally, the "checkerboard" state in $RTe_2$ must be induced by applying pressure[8,9].

Multiple "striped" CDW phases and superstructures in $LaTe_2$ have been experimentally identified via transmission electron microscopy (TEM) measurements[10], encompassing a commensurate CDW wave vector of $q = 0.5a^*$, and an incommensurate CDW wave vectors of $q = (1\text{-}\alpha)a^*$ modulated by defect or substitution, as well as the superstructures derived from the modulation order of Te vacancies. The modulation structure corresponding to $q = 0.5a^*$ was initially attributed to a Fermi surface nesting vector or a superstructure of $LaTe_2$[11], however, subsequent phonon spectrum calculations demonstrated that it arises from phonon softening near the $X$ point ($q \sim 0.5a^*$)[12,13]. Substituting Te with Sb of similar ionic radius enables continuous tunability of the Fermi surface and the incommensurate CDW nesting wave vector $q = (1\text{-}\alpha)a^*$[14]. Collectively, these investigations reveal that CDW instability in $LaTe_2$, particularly in the Sb-substituted crystals $LaTe_{2\text{-}x}Sb_x$, stems from diverse origins, where the synergistic interplay of Fermi surface nesting and electron-phonon coupling gives rise to multiple CDW orders[12]. Nevertheless, certain complex diffraction spot features observed in TEM measurements (e.g., satellite spots or additional wave vectors) cannot be fully elucidated by the aforementioned conclusions[14]. High-precision surface characterization techniques such as scanning tunneling microscopy (STM) are therefore required to uncover the origins of these unaccounted-for structural features. Notably, further experimental studies have uncovered unexpected quantum states and physical properties: for instance, an axial Higgs-type charge density wave amplitude mode sensitive to magnetic fields was discovered in the isostructural compound $RTe_3$[15], and magnetic-field-induced electronic structure modifications lead to highly anisotropic large transverse magnetoresistance (MR), as well as an anomalously large positive MR in $LaTe_3$[16].

In this work, we present a scanning tunneling microscopy (STM) investigation of $LaTe_{1.6}Sb_{0.4}$ crystals with 20% Sb substitution, which have been rarely explored in previous studies. $LaTe_{1.6}Sb_{0.4}$ is a multiple CDW compound at ambient pressure and undergoes a semiconductor-metal-superconductor transition upon application of pressure between 2.5 and 4.6 GPa. Two distinct CDW modulations, $q_1$=0.73$a^*$ and $q_2$=0.5$a^*$, have been observed by transmission electron microscopy (TEM), which were attributed to Fermi surface nesting and electron-phonon coupling, respectively[12]. In the present study, using high-resolution STM topography and fast Fourier transform (FFT) imaging, we observe a striped superstructure with a periodicity of 11 times the lattice constant, as well as three mutually interweaved CDW wave vectors. The superstructure (with wave vector $p=\pm0.27a^*+b^*$) arise from the combined effects of Sb substitution and the formation of covalent-like quasi-bonding (CLQB)[17,18]. The three interweaved CDW vectors : $q_1$=0.73$a^*$ (8/11$a^*$), $q_2$=0.47$a^*$ (5/11$a^*$) and $q_3$=0.27$a^*$ (3/11$a^*$) satisfy the relations $q_1+q_3=a^*$ and $q_2= q_1-q_3$, indicating that they originate from the hole doping and the contribution of 5$p$ orbitals. The wave vector $q_1$ and $q_3$ are theoretically illustrated to be nesting vectors between $p_x$- and $p_y$-bands. Our results provide important insights for further exploring the hole/electron doping effect on CDW materials.

## Materials and Methods

### Sample preparation and STM measurements.

High quality $LaTe_{1.6}Sb_{0.4}$ single crystals are synthesized by chemical vapor transport reactions in quartz tube using $I_2$ as the transporting agent. The starting materials are La ingot (Alfa, 99.99%), Te pieces (Alfa, 99.999%) and Sb grains (Alfa, 99.999%). The high and low temperatures for the reaction are set as 1120K and 1220K, respectively. The prepared crystals are stored in an dry and oxygen-free atmosphere to prevent oxidation reactions. After the growth experiment, the sample was transferred to an UHV chamber with LT-STM (PanScan Freedom, RHK) for the following STM measurements. The $LaTe_{1.6}Sb_{0.4}$ single crystals are exfoliated along the *ab*-plane in the UHV chamber. All STM measurements were performed at ~9 K with chemical etched W tip. The STM tip condition was carefully calibrated on Au(111) surface by measuring the Au(111) Shockley surface state before the STS measurements. Gwyddion software was used for STM data analysis.

**Characterization**

Chemical compositions of $LaTe_{1.6}Sb_{0.4}$ single crystals were determined by Energy Dispersive Spectroscopy (EDS) and Inductively Coupled Plasma (ICP). EDS mapping images were obtained using an ARM-200F (JEOL, Tokyo, Japan). Selected area electron diffraction (SAED) was performed by using a high-resolution transmission electron microscopy (HRTEM, Tecnai F20 super-twin). Electrical resistivity ($\rho$) was measured through the physical property measurement system (PPMS, Quantum Design).

**Theoretical calculations**

In this paper, we investigate the structural and electronic properties of Sb-doped $LaTe_2$, where one out of every four Te atoms is substituted by an Sb atom. Doping-induced structural reconstruction restores the system's symmetry to a tetragonal lattice, with the corresponding space group redefined accordingly. Our computational approach is detailed as follows: Structural optimization, electronic structure, and magnetic properties calculations were performed using the Vienna *Ab* initio Simulation Package (VASP) code, employing the projected augmented-wave (PAW) pseudopotential formalism. The Perdew-Burke-Ernzerhof (PBE) variant of the generalized gradient approximation (GGA) was adopted to model the exchange-correlation functional. A plane-wave cutoff energy of 500 eV was employed. For geometry relaxation and self-consistent field (SCF) calculations, the Brillouin zone was sampled using an 8×8×4 $\Gamma$-centered Monkhorst-Pack *k*-mesh and an 8×8×4 Monkhorst-Pack *k*-mesh (generated with a *k*-spacing of 0.030), respectively. All geometry optimizations were executed via the conjugate gradient method, with force and energy convergence criteria set to 0.02 eV/Å and $1\times10^{-6}$ eV, respectively, to strike a balance between computational accuracy and efficiency. Subsequently, leveraging the wavefunction files from the preceding SCF calculation, we performed two distinct types of non-self-consistent calculations. First, a *Γ*-centered 10×10×6 *k*-point mesh was employed for uniform Brillouin zone sampling, and 2D Fermi surface images of the doped $LaTe_2$ were generated using the IFermi software package. Second, non-self-consistent calculations were conducted in Line-mode along the high-symmetry path *Γ-X-M-Γ-Z-R-A-Z* of the tetragonal Brillouin zone, with 60 *k*-points allocated per path segment. From these calculations, the band structure and projected band structure of the doped $LaTe_2$ system were derived.

## Results and Discussion

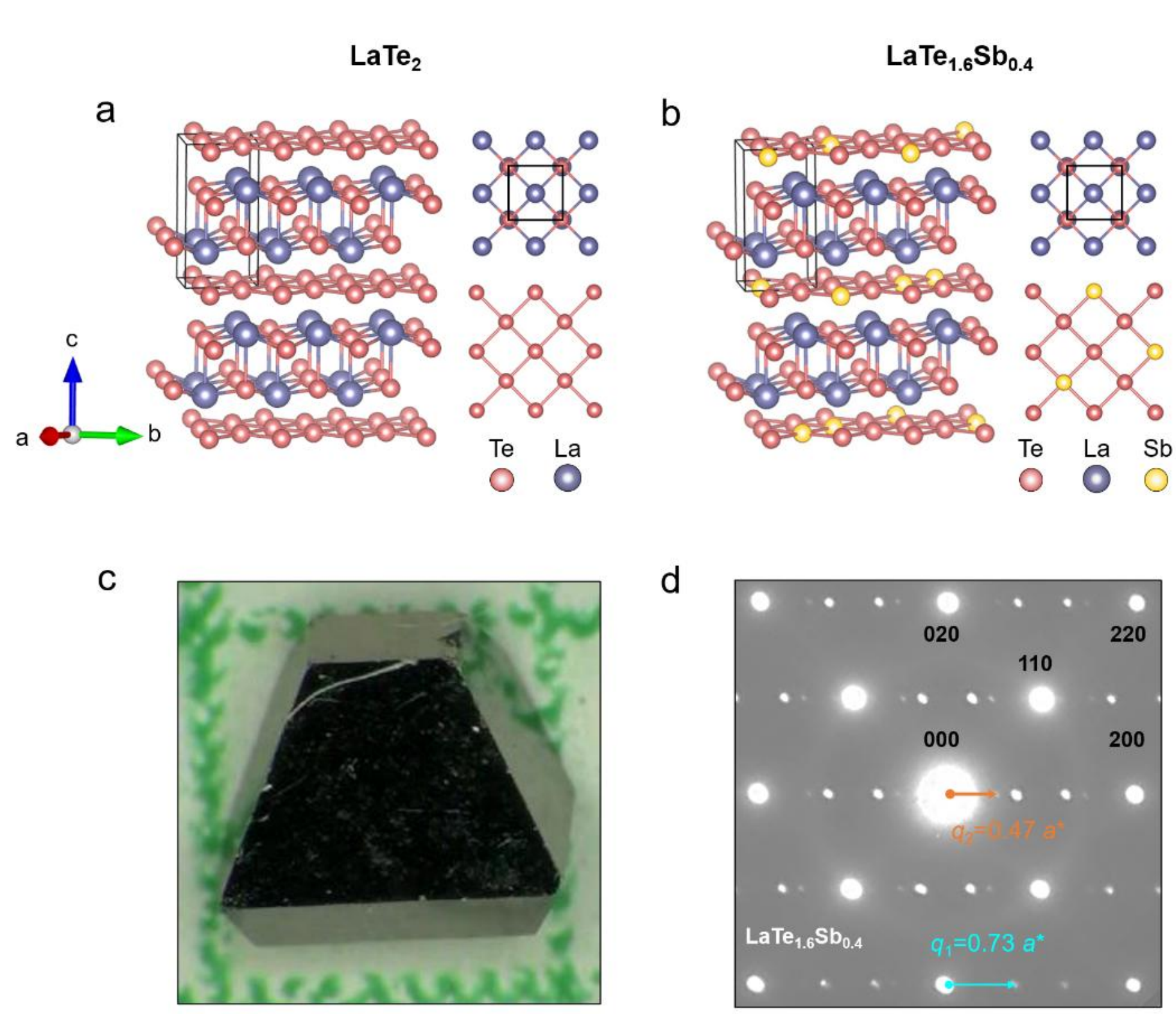

**Fig. 1. Atomic structural of $LaTe_2$/$LaTe_{1.6}Sb_{0.4}$, and charge density waves in $LaTe_{1.6}Sb_{0.4}$.** (a) Left panel: crystal structure of $LaTe_2$. Right panel: top view of and Te-Te layer. (b) Left panel: crystal structure of $LaTe_{1.6}Sb_{0.4}$. Right panel: top view of La-Te and Te-Te (Sb) layer. (c) Optical photograph of $LaTe_{1.6}Sb_{0.4}$ single crystal. The length of the grid lines is 2 mm. (d) Selected area electron diffraction of $LaTe_{1.6}Sb_{0.4}$ performed at room temperature along [001] axis. Strong (*hk0*) reflections with $h$+$k$ even correspond to the sublattice. Two CDW wave vectors $q_1 = 0.73\ a^*$ and $q_2 = 0.47\ a^*$ can be distinguished, respectively.

Figure 1a shows a schematic diagram of the $LaTe_2$ crystal, which adopts the $Cu_2Sb$-type structure with (P4/*nmm*) symmetry. The black solid line denotes the unit cell, with its *c*-axis oriented vertically. In this two-dimensional layered $LaTe_2$ crystal, La-Te layers alternate with Te-Te square nets, which display fourfold rotational symmetry within the *ab*-plane. Based on previous transmission electron microscope (TEM) investigations[10], the *in-plane* tetragonal lattice parameter of $LaTe_2$ was measured to be $a = 4.55 \pm 0.02$ Å. The $LaTe_{1.6}Sb_{0.4}$ crystal possesses a structure similar to that of $LaTe_2$ (Fig. 1b). Its lattice constants are $a = 4.50$ Å and $b = 4.51$ Å[12]. For Te-Te square sheets, the wide formation range of $RTe^{2-}$ compounds gives rise to a strong tendency toward substantial Te vacancies on the square planar Te sites[19-21].

Accordingly, substituting Te atoms in the Te square sheets with Sb atoms of comparable atomic radius has been established as an effective strategy to tune the electronic structure of $LaTe_2$ [14].

The 20% Sb-substituted $LaTe_{1.6}Sb_{0.4}$ crystal has a black, plate-like morphology, as shown in Fig. 1c, with dimensions of 1.8 mm × 1.2 mm × 0.5 mm (length × width × thickness). The crystal was fabricated by chemical vapor transport reactions, with atomic ratios of La: Te: Sb are 1: 1.56: 0.37 as reported by the previous ICP measurements[12] confirmed the atomic ratio of La:Te:Sb to be 1:1.56:0.37. As shown in Supplementary information Fig. S1, the $LaTe_{1.6}Sb_{0.4}$ single crystal exhibits semiconducting behavior. The coexistence of multiple CDW phases in $LaTe_{1.6}Sb_{0.4}$ is verified by selected-area electron diffraction patterns acquired along the [001] zone axis. As presented in Fig. 1d, two distinct sets of wave vectors are clearly observed: $q_1$= 0.73 $a^*$ (marked by cyan arrows) and $q_1$= 0.5 $a^*$ (marked by orange arrows), in good agreement with the previous report[12].

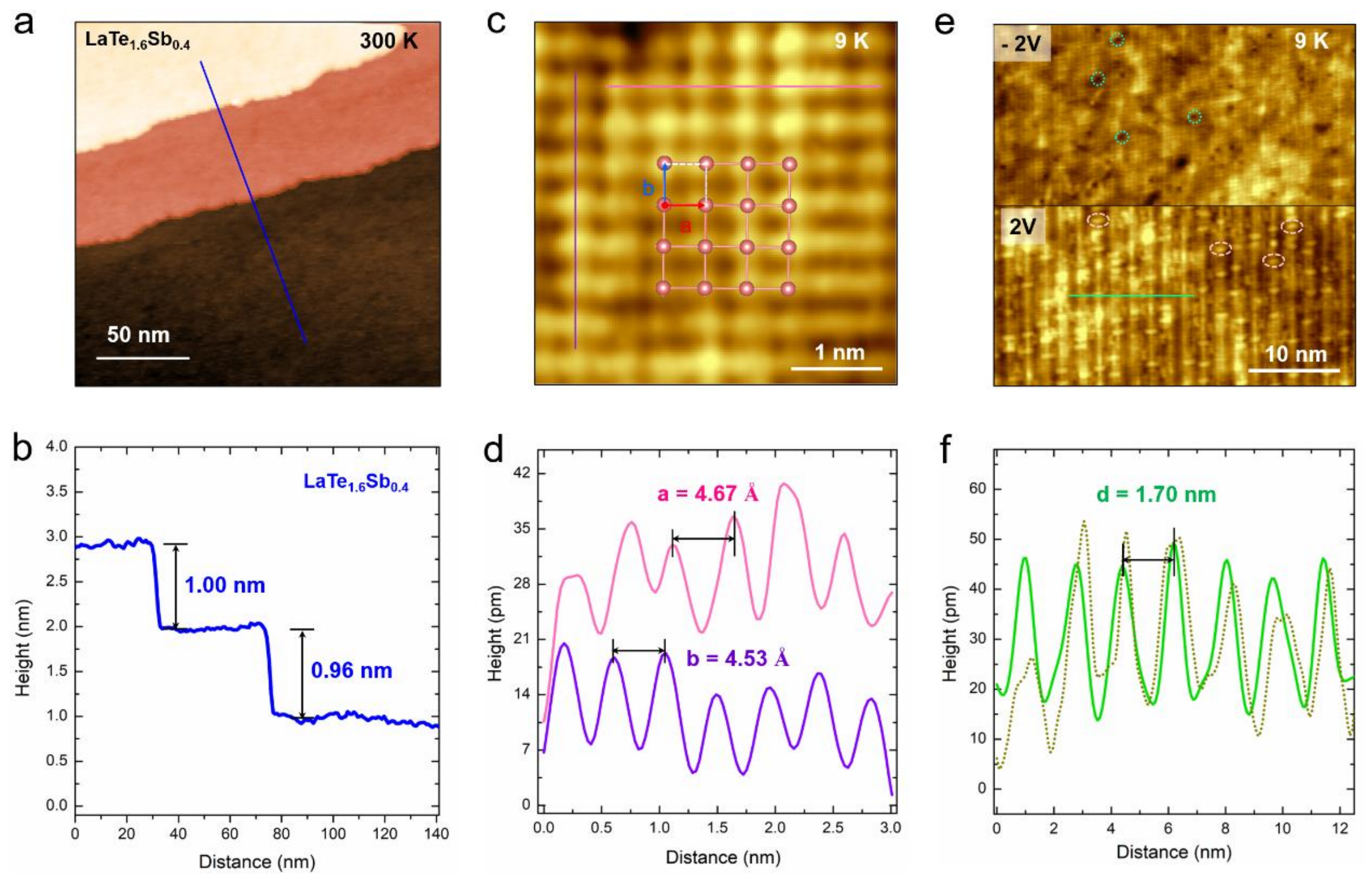

**Fig. 2. Atomic resolution of Te-Te (Sb) cleavage surface in $LaTe_{1.6}Sb_{0.4}$ and its bias-dependent STM topography acquired at 9 K.** (a) Large-scale STM topography of $LaTe_{1.6}Sb_{0.4}$ ($V$ = -1.7 V, $I$ = 120 pA) with the edges performed at 300 K. (b) The line profiles show a single step height of approximately 1nm in $LaTe_{1.6}Sb_{0.4}$. (c) Atomic-resolution STM image of Te-Te (Sb) cleavage surfaces obtained at 9 K in $LaTe_{1.6}Sb_{0.4}$, ($V$ = -1.8 V, $I$ = 100 pA). (d) The height line profiles along the $a$-(pink line) and $b$-axis (purple line) in (c), respectively. The lattice constant of $LaTe_{1.6}Sb_{0.4}$ is measured to be 4.67 Å ($a$) and 4.53 Å ($b$). (e)

Comparison STM topography image in-*situ* at occupied (the up panel) and unoccupied states (the down panel) of the cleaved Te-Te (Sb) surface in the same crystal, ($V$ = ± 2.0 V, $I$ = 50 pA). The blue circle and pink oval circle in Fig.2(d) mark the positions of Te defects and Sb substitutions, respectively. (f) The height line profiles of filtered (green solid line) and unfiltered ( olive dotted line) lattice characteristics along the green line in Fig. 2(e), which exhibits a periodic modulation along the $a$-axis of approximately 1.70 nm.

Figure 2a presents a large-scale STM topographic image of the cleaved surface of $LaTe_{1.6}Sb_{0.4}$ at 300 K. The surface exhibits irregular, line-shaped step edges and a relatively rough terrace morphology. There are no CDW characteristics can be observed in the zoom-in STM topography performed at this temperature (Supplementary information Fig. S2). As displayed in Fig. 2b, the average single-layer step height of $LaTe_{1.6}Sb_{0.4}$ single crystals is measured to be approximately 1 nm.

Figure 2c shows a zoom-in high-resolution atomic image of the cleaved surface of $LaTe_{1.6}Sb_{0.4}$ at 9 K. All atoms exhibit uniform brightness, ruling out the possibility that this surface corresponds to the La-Te plane. The Te-Te (Sb) cleavage surface of $LaTe_{1.6}Sb_{0.4}$ displays approximate $C_4$ symmetry, with measured lattice constants of 4.67 Å ($a$-axis) and 4.53 Å ($b$-axis). A slight discrepancy exists between the experimental values and the theoretical values (a = 4.50 Å, $b$ = 4.51 Å) [12,14], which can be attributed to *in-plane* strain or other surface effects. Figure 2d presents the line profiles along the $a$- and $b$-axes, revealing the surface corrugation of the Te-Te (Sb) cleavage plane induced by surface strain and Sb substitution.

Figure 2e shows distinct differences between the occupied (upper panel) and unoccupied states (lower panel) of the Te-Te (Sb) cleavage surface measured *in situ*. In the occupied state, only the lattice features of the Te-Te (Sb) square net are resolved. The cyan circles mark Te atom vacancies, which appear as dark spots on the cleavage surface. In contrast to the uniform lattice observed in the occupied states, prominent striped modulation features along the $a$-axis are clearly visible in the unoccupied states. In this state, Sb substitution becomes discernible, appearing as numerous bright bars (marked by pink ellipses) along the $b$-axis. As shown in Fig. 2f, the stripe-shaped modulation in the unoccupied state exhibits a period of approximately 1.70 nm. Line profiles of both the raw (green) and filtered lattice signals (olive) reveal the same incommensurate modulation period (1.70 nm≈3.64$a$).

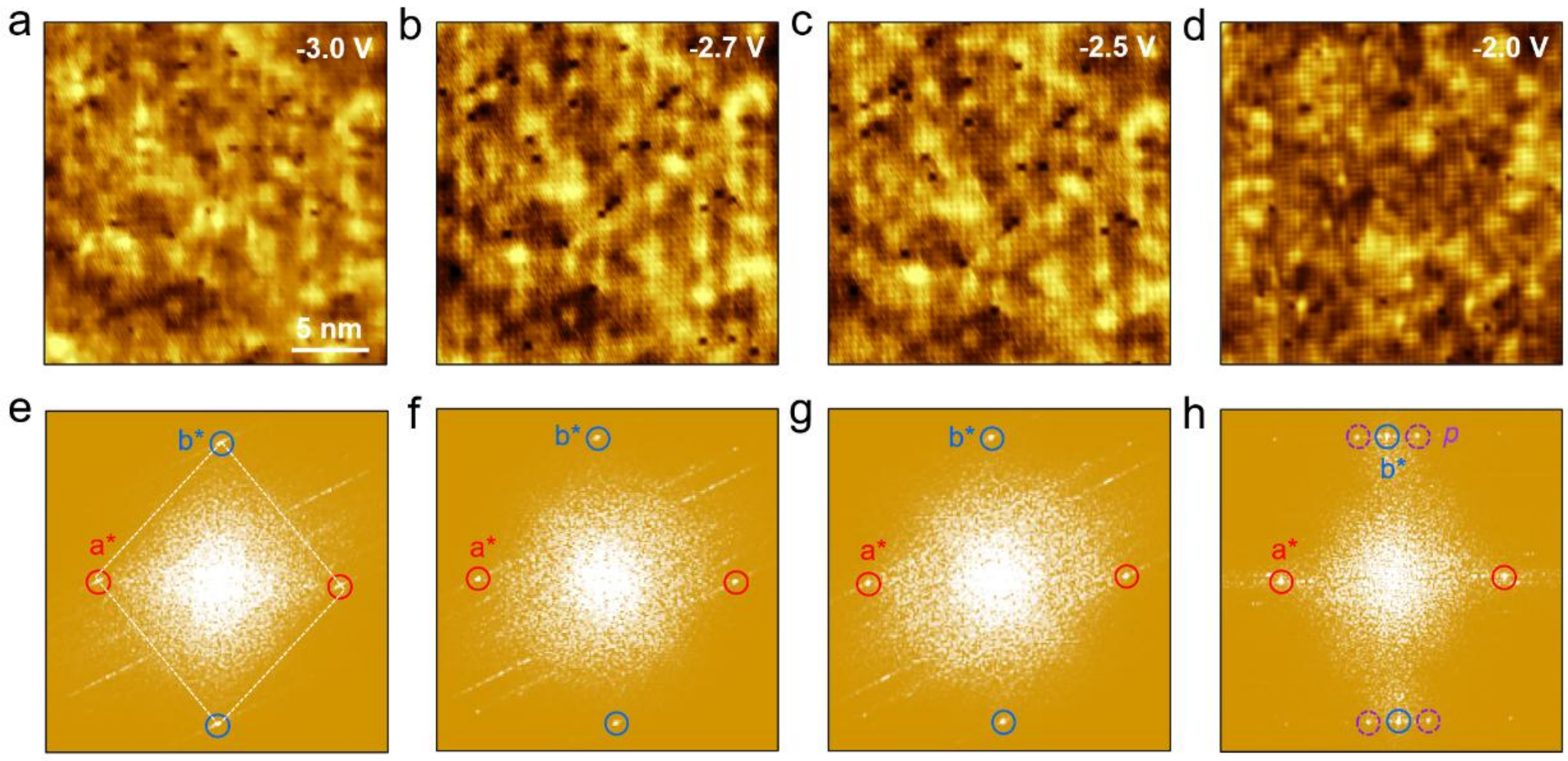

**Fig. 3. Atomic-scale STM measurements of the $LaTe_{1.6}Sb_{0.4}$ in occupied state.** (a-d) STM topography of Te-Te (Sb) cleavage surface obtained at 9K from -3.0 V to -2.0 V ($I$ = 100 pA), no ordered structure can be observed except for the Te-Te (Sb) lattice until -2.0V. (e-h) Fast Fourier transform (FFT) patterns correspond to (a-d), respectively. Only a pair of satellite spots of $b^*$ are resolved in Fig. 3(g), as marked by purple circles $p = 0.27a^* \pm b^*$ in Fig. 3(h).

To investigate the topographic differences between the occupied and unoccupied states of the Te-Te (Sb) cleavage surface, zoom-in STM images and corresponding FFT patterns are presented in Figs. 3 and 4. Figures 3(a)-3(d) show magnified images acquired in the occupied states, in which the Te-Te (Sb) square-net lattice features can be clearly resolved in each panel. From -3.0 V to -2.0 V, faint strip-shaped textures can be barely distinguished in real space (see Supplementary Information Fig. S3).

Figure 3(e)-3(h) display the corresponding fast Fourier transform (FFT) patterns of Figs. 3(a)-3(d). In Figs. 3(e)-3(g), only Bragg spots (marked by red and blue circles) are resolved under occupied states. However, at -2V, due to the shortened distance between the STM tip and sample, a set of satellite spots[22-24] with $p = 0.27a^* \pm b^*$ emerges on both sides of $b^*$ (indicated by purple circles in Fig. 3(h)). There are no characteristics can be resolved in the 4 times magnified zoom-in FFT patterns as shown in Supplementary information Fig. S4. Given the stable presence of wave vector $p$ over a higher energy range (from to), we infer that it likely corresponds to a striped superstructure. According to a previous study, Yang *et al*. employed high-angle annular dark-field scanning transmission electron microscopy (HAADF-STEM) to observe a superstructure induced by periodic Te defects in $LaTe_{2-\delta}$, with a modulation wave vector of $q = 0.2a^* + 0.6b^*$[10]. Nevertheless, the difference between $p$ and $q$ is substantial, making it difficult to form a four-fold symmetric superstructure. Furthermore, Sb substitution reduces Te vacancies within the Te-Te square lattice, suggesting that the wave vector $p$ is unlikely to originate from periodic Te vacancies. Another possibility is that, in CDW systems, satellite spots typically indicate a CDW superlattice. For instance, a continuous evolution of the CDW lattice has been reported in 1T-$TaS_2$, driven by strong coupling between the NC-CDW state and the lattice [24,25]. Therefore, the satellite spots ($p$) are possibly closer to the striped CDW superlattice.

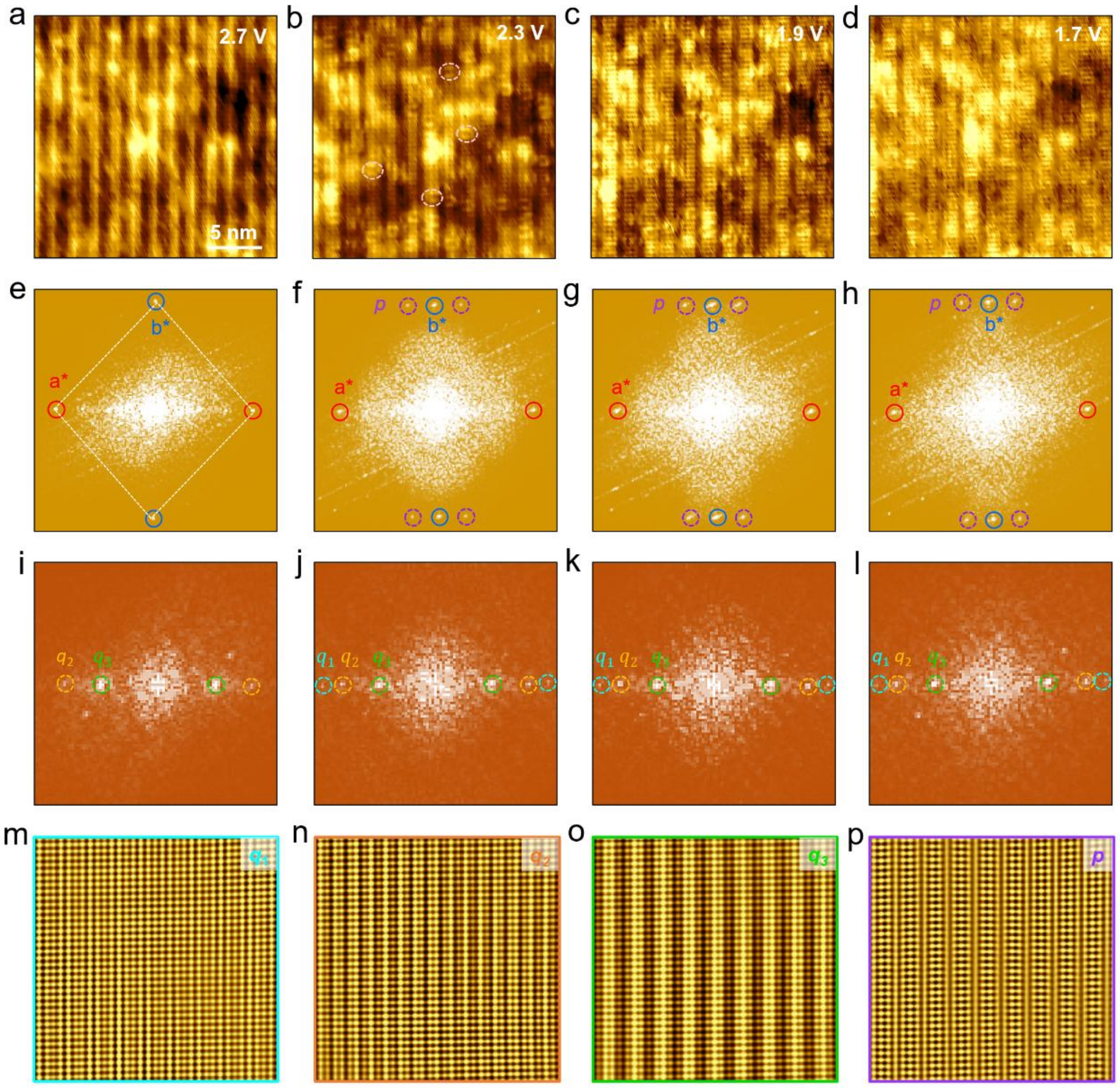

**Fig. 4. Charge density waves and superstructure in $LaTe_{1.6}Sb_{0.4}$ at unoccupied state.** (a-d) STM topography of Te-Te (Sb) cleavage surface obtained at 9K from 2.7 V to 1.7 V ($I$ = 100 pA), any of them exhibit stripe modulation characteristics along the $a$-axis, which are significantly different from those at occupied states. The the Sb substitutions in Fig. 4(b) are marked by pink oval circles; it is noted that the electronic states at 2.3V are different from that at 2V (see Fig. 2(c)). (e-h) are the corresponding FFT patterns of Fig. 4(a-d). The characteristics of the superstructure $p = 0.27\ a^* + b^*$ (purple circles) emerge from 2.5V to 1.7V, which can preserve in a quite large energy scale (until -2.2V, see supplementary information Fig. S3. (i-j) are zoom-in FFT patterns (4 times magnified) of Fig. 4(e-h), which represent the characteristics of three charge density waves $q_1 = 0.73a^*$ ($8/11a^*$), $q_2 = 0.47\ a^*$ ($5/11a^*$) and $q_3 = 0.27\ a^*$ ($3/11a^*$). $q_1$, $q_2$ and $q_3$ are marked in cyan, orange and green circles, respectively. (m-p) Inverse Fourier transform (IFFT) images of CDW $q_1$, $q_2$, $q_3$ and superstructure $p$, from which distinct periodic modulation can be observed.

We next investigated the electronic states of the Te-Te (Sb) cleavage surface in the unoccupied states by examining topographies acquired at different bias voltages (from 1.7 V to 2.7 V) and their corresponding FFT images, as shown in Fig. 4 and Supplementary information Fig. S5. Figure 4(a) presents the topography image obtained at 2.7 V, which exhibits a striped modulation along the $a$-axis markedly different from those observed in the occupied states (Fig. 3). At 2.3 V, both the stripe-like modulation and Sb substitutions (highlighted by pink elliptical circles) are clearly visible, as shown in Fig. 4(b). The Te-Te (Sb) square-net lattice can also be resolved at this bias. Upon further narrowing the energy window (1.9 V, Fig. 4(c)), clear lattice textures become distinguishable within the modulated stripes, while the features associated with Sb substitutions gradually become blurred. When the bias voltage is reduced to 1.7 V (Fig. 4(d)), the circular ring-like features arising from Sb substitutions disappear completely, whereas the stripe modulation and lattice texture remain preserved. Notably, the stripe modulation along the $a$-axis in the unoccupied states does not appear as a single periodic structure, but rather as a superposition of multiple periods.

Figure 4(e) shows the FFT of the topography acquired at 2.7 V, which reveals only the Bragg spots corresponding to the Te-Te (Sb) square-net lattice. In contrast, the FFTs of the topographies obtained in the unoccupied states from 1.7 V to 2.3 V exhibit additional satellite spots (denoted as $p$), as shown in Figs. 4(f)-4(h). These superstructure features persist over a wide energy range, from -2 V to 2.3 V.

To further analyze the periods associated with the stripe modulations observed in Figs. 4(a)-4(d), the central regions of the FFT images shown in Figs. 4(e)-4(h) have been magnified fourfold. Three distinct sets of CDW wave vectors with different periods can be clearly resolved, labeled from the outermost to the innermost as $q_1$ = 0.73 $a^*$ (cyan), $q_2$= 0.47 $a^*$ (orange), and $q_3$ = 0.27 $a^*$ (green). Among these, wave vector $q_1$ appears only within a narrow energy range from 2.3 V to 1.7 V, whereas $q_2$ and $q_3$ remain stable across the broader energy range of 2.7 V to 1.7 V. These two CDW wave vectors are in good agreement with TEM data reported for $LaTe_{1.6}Sb_{0.4}$ in previous literature, specifically $q_1$ = 0.73 $a^*$ and $q_2$ = 0.50 $a^*$ [12]. The $q_2$ = 0.50 $a^*$ wave vector has been attributed to Fermi surface nesting in the parent $LaTe_2$ phase, as confirmed by ARPES measurements[11]. The other wave vector, $q_1$ = 0.73 $a^*$, corresponds to a progressively increasing nesting vector as electrons are removed from the system upon Sb substitution[14]. Inverse Fourier transform images (see Supplementary Information Fig. S6 for more details) corresponding to the CDW wave vectors $q_1$, $q_2$, $q_3$ and the superstructure wave vector $p$ are presented in Figs. 4(m)-4(p), respectively, all of which exhibit stripe-like modulation patterns in real space.

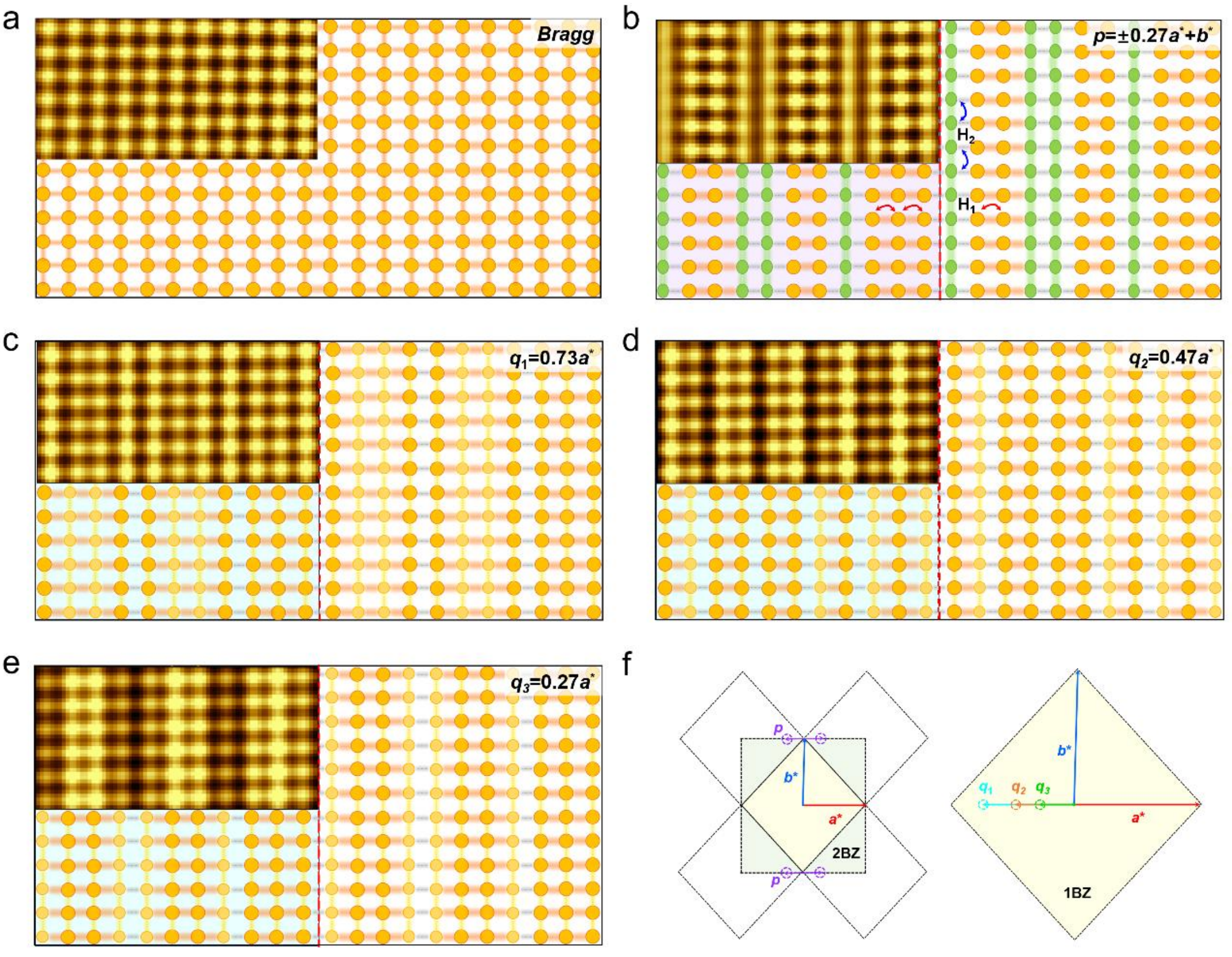

**Fig. 5. Schematic diagram of the charge density waves and superstructure of $LaTe_{1.6}Sb_{0.4}$ in real space and reciprocal space.** (a) Atomic resolution of Te-Te (Sb) cleavage surface performing inverse Fourier transform (IFFT) on STM experimental images (the top left panel), filtered from Fig. 4(c). The $C_4$ symmetric

structure is completely consistent with the atomic structure in the bottom right panel. (b-e) show the evolution from the IFFT of the superstructure and CDW of $LaTe_{1.6}Sb_{0.4}$ to the atomic structures (from top left panel to bottom right panel), respectively. All modulation structures have a lattice constant period of 11 times (along the *a*-axis, marked by purple or green shadow). (b) Schematic model of the electronic superstructure $p$=0.27$a^*$+$b^*$. The hopping along the *a*-axis ($H_1$ in red) and *b*-axis ($H_2$ in blue) between the *p*-orbitals of the nearest neighbor Te atoms give rise to the orange and green electronic state structures in the Te square lattice, respectively. (c) CDW with $q_1$=0.73$a^*$ (8/11$a^*$) wave vector. (d) CDW with $q_2$=0.47$a^*$ (5/11$a^*$) wave vector. (e) CDW with $q_3$=0.27$a^*$ (3/11$a^*$) wave vector. The orange and yellow circles represent atoms with brighter and darker contrast, respectively. (f) The distribution of the electronic superstructure and CDW wave vectors in reciprocal space. The yellow and green areas represent the first and second Brillouin zones of $LaTe_{1.6}Sb_{0.4}$ in the *q*-space, respectively.

To elucidate the nature of the wave vectors $p$, $q_1$, $q_2$, and $q_3$, as well as their corresponding real-space modulations, we constructed schematic diagrams based on the Te-Te square net lattice, as shown in Fig. 5. In Fig. 5(a), each Te atom forms four covalent bonds with its nearest neighbors, resulting in a square-net lattice with $C_4$ symmetry. Figure 5(b) illustrates the superstructure arising from the satellite spots $p = 0.27a^* + b^*$. The inverse fast Fourier transform (IFFT) corresponding to $p$, combined with the lattice characteristics, reveals a 11-*a*-superlattice along the *a*-axis, highlighted by the purple shaded region. This uniaxial periodic modulation arises from alternating arrangements of trimers and dimers (marked by orange dots) along the *a*-axis and single-atom chains (marked by green elliptical dots) along the *b*-axis. Each 11-*a*-superlattice contains two tetramers, three chains along the *b*-axis, and one trimer (see supplementary information Fig. S7). According to previous theoretical studies, in rare earth tellurides, the dominant interaction among nearest-neighbor *p* orbitals is hopping rather than hybridization[8]. So, the hopping amplitude $H_1$ (red curve path), along the extended direction of the given *p* orbital (*a*-axis), has been theoretically demonstrated to be much larger than the hopping amplitude $H_2$ (blue curve path) perpendicular to the extended direction (*b*-axis), due to the highly anisotropic profile of the *p* orbital electron wave function[8]. As a result, the trimer and dimer in Fig. 5(b) have a brighter contrast compared to the single atom chains. Therefore, we believe that the stripe modulation shown in Fig. 5(b) is closer to an electronic superstructure.

The charge density waves corresponding to $q_1$, $q_2$, and $q_3$ share the same modulation period along the *a*-axis, with a superperiod of 11 times the lattice constant (indicated by green shaded regions in Figs. 5(c)-5(e)). The distinctions among them arise solely from differences in the substructures within a CDW-period. Inverse fast Fourier transform (IFFT) analysis effectively reveals the real-space difference of these three wave vectors: (1) The CDW with wave vector $q_1 = 0.73\ a^*$ (i.e., $8/11\ a^*$) comprises eight columns of bright atoms (orange circles) and three columns of dark atoms (yellow circles) within one modulation period, as shown in Fig. 5(c). (2) The CDW with $q_2 = 0.47\ a^*$ (i.e., $5/11\ a^*$) consists of five columns of bright atoms and six columns of dark atoms per period (Fig. 5(d)). (3) The CDW with $q_3 = 0.27\ a^*$ (i.e., $3/11\ a^*$) contains three columns of bright atoms and eight columns of dark atoms per period (Fig. 5(e)). Interestingly, $q_1 = 8/11\ a^*$ and $q_3 = 3/11\ a^*$ satisfy the relation $q_1 = a^* - q_3$, while $q_2 = 5/11\ a^*$ can be expressed as $q_2 = q_1 - q_3$. This suggests that $q_1$, $q_2$, and $q_3$ may represent three interweaved components of an incommensurate CDW.

Figure 5(f) presents a schematic diagram illustrating the distribution of the aforementioned wave vectors in reciprocal space. In the left panel of Fig. 5(f), the wave vectors of the electronic superstructure, denoted as $p = 0.27a^* + b^*$ (shown in purple), are located in the second Brillouin zone (2BZ), acting as a pair of satellite spots associated with the Bragg vector $b^*$ (blue). In the right panel, the CDW wave vectors $q_1$ (cyan), $q_2$(orange), and $q_3$ (green) are situated in the first Brillouin zone (1BZ) and aligned along the $a^*$-axis (red). Notably, the electronic superstructure and $q_3$ share the same modulation period of $3/11a^*$, which corresponds to a real-space period of approximately 1.71 nm (given by $(1/0.27) \times a$). This value is in excellent agreement with the stripe-like periodic modulation observed in Fig. 2(e), as well as those shown in Figs. 3(a)-3(d) and Figs. 4(a)-4(d).

To investigate the origin of the superstructure, the period of the 11-$a$-superlattice superstructure was measured to be 5.09 nm (see Supplementary Information Fig. S8 for more details). The average distance between atomic centers within the superstructure is therefore 5.09 nm/11 $\approx$ 4.63 Å. Subtracting the ionic radius of a single Te ion in the square sheet (1.34 Å) yields an average interatomic spacing of approximately 3.29 Å. In a pristine Te-Te square net, the average spacing between Te atoms is approximately 3.1 Å - a value that lies precisely at the boundary between normal bonding and non-bonding. This unstable arrangement in the Te-Te (Sb) square net is energetically mitigated by the formation of covalent-like quasi-bonding (CLQB) among Te atoms on neighboring diagonals[26,27] which gives rise to a series of incommensurate charge density waves[26-29]. In the present case, where the square sheet contains 20% Sb substitution ($LaTe_{1.6}Sb_{0.4}$), the atomic spacing along the *a*-axis is slightly larger than 3.1 Å, yet remains below the 3.4 Å threshold required for CLQB formation[26]. We therefore propose that CLQB initially forms within the Te-Te (Sb) sheet, leading to the emergence of the superstructure *p*, which manifests as a stripe modulation with a real-space period of 1.71 nm.

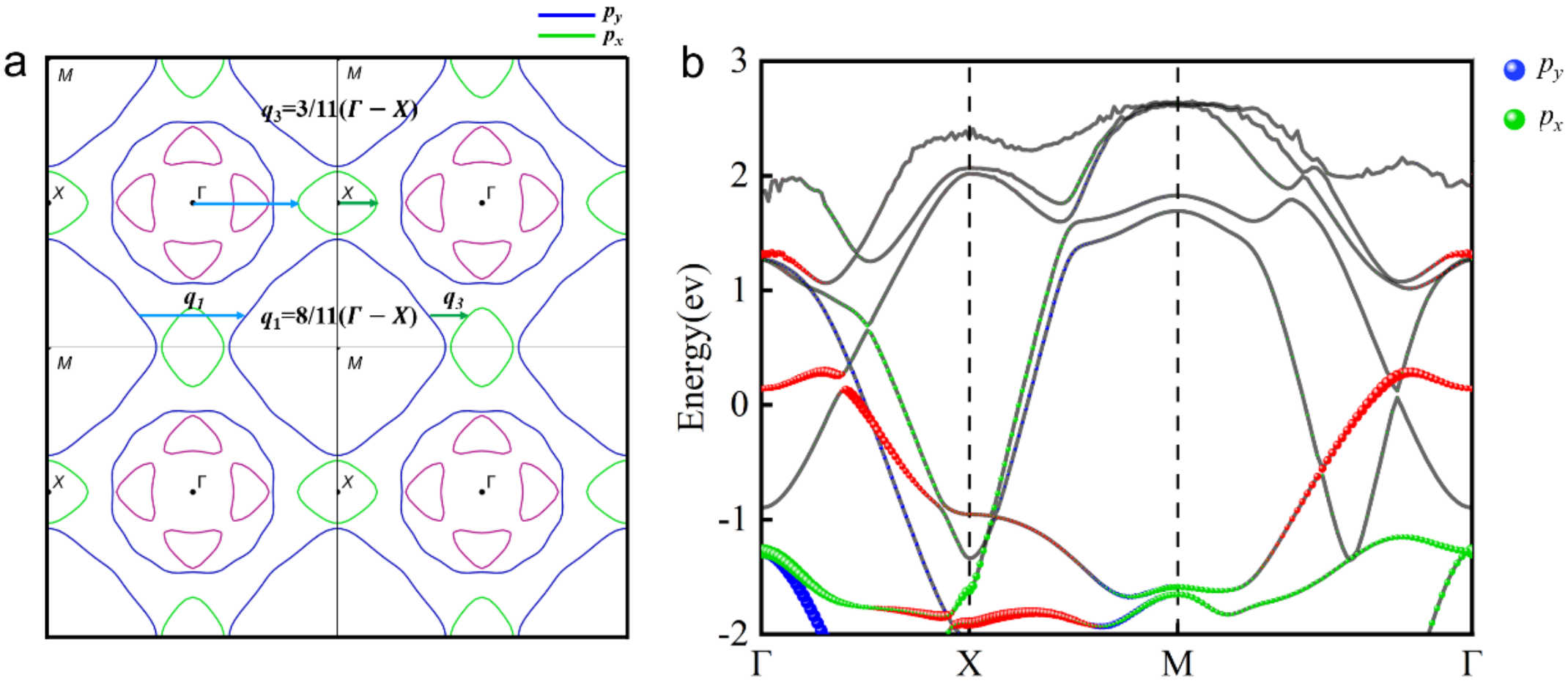

**Fig. 6. Theoretical electronic structure of the $LaTe_{1.5}Sb_{0.5}$ crystal.** (a) 2D-Fermi surface in the 2D Brillouin zone ($k_z$ =0), with the cyan and green arrows indicate two typical nesting vectors ($q_1$=8/11($\Gamma - X$) and $q_3$=3/11($\Gamma - X$)) relevant with the CDW, respectively. (b) The projected electronic band structures of Te atoms in bulk $LaTe_{1.6}Sb_{0.4}$. The *p* orbitals of Te atoms in (b) are mapped with different colors: Te-$p_x$, green; Te-$p_y$, blue.

We have performed detailed electronic structure calculations for $LaTe_{1.5}Sb_{0.5}$, with the results presented in Fig. 6. From the two-dimensional Fermi surface, two nesting vectors $q_1$=8/11($\Gamma - X$) (marked in blue arrows) and $q_3$=3/11($\Gamma - X$) (marked in green arrows) can be clearly identified. These are in good agreement with the experimentally observed CDW wave vectors. The calculated 2D-Fermi surface in the 2D Brillouin zone ($k_z$=0) of Te-Te square net primary is composed of 5$p_x$ and 5$p_y$ orbitals. The two nesting vector: $q_1$=8/11($\Gamma - X$)(marked in blue arrows) connecting the $p_y$-$p_y$-derived bands and $q_3$=3/11($\Gamma - X$)(marked in green arrows) connecting the original $p_y$-bands and folded $p_x$-bands, are consistent with the experimental CDW wave vectors (Fig. 6(a)), respectively. Furthermore, the projected electronic band structures of Te atoms in bulk $LaTe_{1.5}Sb_{0.5}$ reveal that the CDW primarily derive from the contribution of 5$p$ orbitals, as illustrated in Fig. 6(b).

**Conclusion**

In summary, the surface electronic structure of the layered compounds $LaTe_{2-x}Sb_x$ (x = 0 and 0.4) was investigated using scanning tunneling microscopy (STM) at 9 K. In the Te-Te (Sb) square sheets of $LaTe_{1.6}Sb_{0.4}$, three charge density waves with distinct wave vectors and a new superstructure were observed, each appearing within different energy ranges. Inverse fast Fourier transform (IFFT) analysis revealed that the superstructure exhibits a stripe modulation along the $a$-axis corresponding to a 11-$a$-superlattice. This superstructure arises from the partial electron deficiency in the square sheets induced by Sb substitution and the formation of covalent-like quasi-bonding (CLQB). Fundamentally, the substructure within each period is governed by hopping effects dictated by the anisotropy of the $p$-orbital electron wave function, consistent with previous theoretical predictions[8]. Associated with this superstructure are three CDW wave vectors: $q_1 = 0.73\ a^*$ (8/11 $a^*$), $q_2 = 0.47\ a^*$ (5/11 $a^*$), and $q_3 = 0.27\ a^*$ (3/11 $a^*$). They satisfy the relations $q_1 = a - q_3$ and $q_2 = q_1 - q_3$, which are interweaved CDW induced by hole doping. The two nesting wave vector: $q_1$ and $q_3$ connect the $p_x$- and $p_y$- bands, as well as their derivative bands, as illustrated in the calculated 2D-Fermi surface. Further spectroscopic analysis, angle-resolved photoemission spectroscopy (ARPES), and temperature-dependent studies would be valuable to elucidate these phenomena in detail.

**Acknowledgments:** This project is supported by the National Natural Science Foundation of China (NSFC) (Grant Nos. 12504213 and 52302010), Natural Science Foundation of Inner Mongolia Department of Science and Technology Autonomous Region Youth Fund (No. 2024QN01010), High level talent introduction and research funding support of Department of Human Resources and Social Security of Inner Mongolia Autonomous Region (No. 12000-150422225). The Postdoctoral Fellowship Program of CPSF under Grant Number. (GZC20250111, 2025M770160); This work was supported by the Synergetic Extreme Condition User Facility (SECUF, https://cstr.cn/31123.02.SECUF) ， including Low-temperature in-situ STM-ARPES experimental station, and sample preselection and characterization station.